\documentclass[12pt]{article}  
\usepackage{graphicx}
\newcommand{\vv}{\vspace*{1.5ex}}   
                    
     \newcommand{\hhh}{\hspace*{9mm}}
\newcommand{\no}{\noindent}
 \newcommand{\bc}{\begin{center}}
 \newcommand{\ec}{\end{center}}
                   \newcommand{\bfr}{\begin{flushright}}
                   \newcommand{\efr}{\end{flushright}}
   \newcommand{\ii}{\item}
     \newcommand{\be}{\begin{enumerate}}
     \newcommand{\ee}{\end{enumerate}}
        \newcommand{\bi}{\begin{itemize}}
        \newcommand{\ei}{\end{itemize}}
            \newcommand{\bd}{\begin{description}}
            \newcommand{\ed}{\end{description}}
                \newcommand{\beq}{\begin{equation}}
                \newcommand{\eeq}{\end{equation}}
                  \newcommand{\bea}{\begin{eqnarray}}
                  \newcommand{\eea}{\end{eqnarray}}

      \newcommand{\bfi}{\begin{figure}}
      \newcommand{\efi}{\end{figure}}
\newcommand{\bay}{\begin{array}{l}}
\newcommand{\eay}{\end{array}}
            \newcommand{\dd}{\mbox{d}}

    \newcommand{\pa}{\partial}
    \newcommand{\del}{\delta}
    \newcommand{\Del}{\Delta}
    \newcommand{\ka}{\kappa}

    \newcommand{\eps}{\epsilon}
    \newcommand{\ga}{\gamma}
    \newcommand{\Ga}{\Gamma}

\begin{document}  

\begin{center}
 {\Large {\sf   Theory of Sorption Hysteresis in Nanoporous Solids: \\[1.7mm]
                I. Snap-Through Instabilities
}}    \\[7mm]

{\sc Zden\v ek P. Ba\v zant\footnote{ McCormick Institute Professor and
W.P. Murphy Professor of Civil Engineering and Materials Science,
Northwestern University, 2145 Sheridan Road, CEE/A135, Evanston, Illinois
60208; z-bazant@northwestern.edu (corresponding author).} and
Martin Z. Bazant\footnote{Associate Professor of Chemical Engineering and
Mathematics, Massachusetts Institute of Technology, Cambridge MA 02139.}}

\vskip 2mm August 22, 2011  

\end{center} \vskip 5mm   

\noindent {\bf Abstract:}\, {\sf The sorption-desorption hysteresis
observed in many nanoporous solids, at vapor pressures low enough for the
the liquid (capillary) phase of the adsorbate to be absent, has long been
vaguely attributed to changes in the nanopore structure, but no
mathematically consistent explanation has been presented. The present work
takes an analytical approach to account for discrete molecular forces in
the nanopore fluid and proposes two related mechanisms that can explain the
hysteresis at low vapor pressure without assuming any change in the
nanopore structure. The first mechanism, presented in Part I, consists of a
series of snap-through instabilities during the filling or emptying of
non-uniform nanopores or nanoscale asperities. The instabilities are caused
by non-uniqueness in the misfit disjoining pressures engendered by a
difference between the nanopore width and an integer multiple of the
thickness of a monomolecular adsorption layer. The second mechanism,
presented in Part II, consists of molecular coalescence within a partially
filled surface, nanopore or nanopore network. This general thermodynamic
instability is driven by attractive intermolecular forces within the
adsorbate and forms the basis to develop a unified theory of both
mechanisms. The ultimate goals of the theory are to predict the fluid
transport in nanoporous solids from microscopic first principles, and 
to determine the pore size distribution and internal surface area from
sorption tests.
 }

\subsection*{Introduction}

The sorption isotherm, characterizing the isothermal dependence of the
adsorbate mass content on the relative vapor pressure at thermodynamic
equilibrium, is a basic characteristic of adsorbent porous solids. It is
important for estimating the internal pore surface of hydrated Portland
cement paste and other materials. It represents the essential input for
solutions of the diffusion equation for drying and wetting of concrete, for
calculations of the release of methane from coal deposits and rock masses,
for the analysis of sequestration of carbon dioxide in rock formations,
etc. Its measurements provide vital information for determining the
internal surface of nanoporous solids \cite[e.g.]{PowBro46, AdoSet96,
Jen00, AdoSet-02, EspFra06, baroghel2007}.

\begin{figure*}
\vspace{-.2in}
\begin{center}
\includegraphics[width=6.5in]{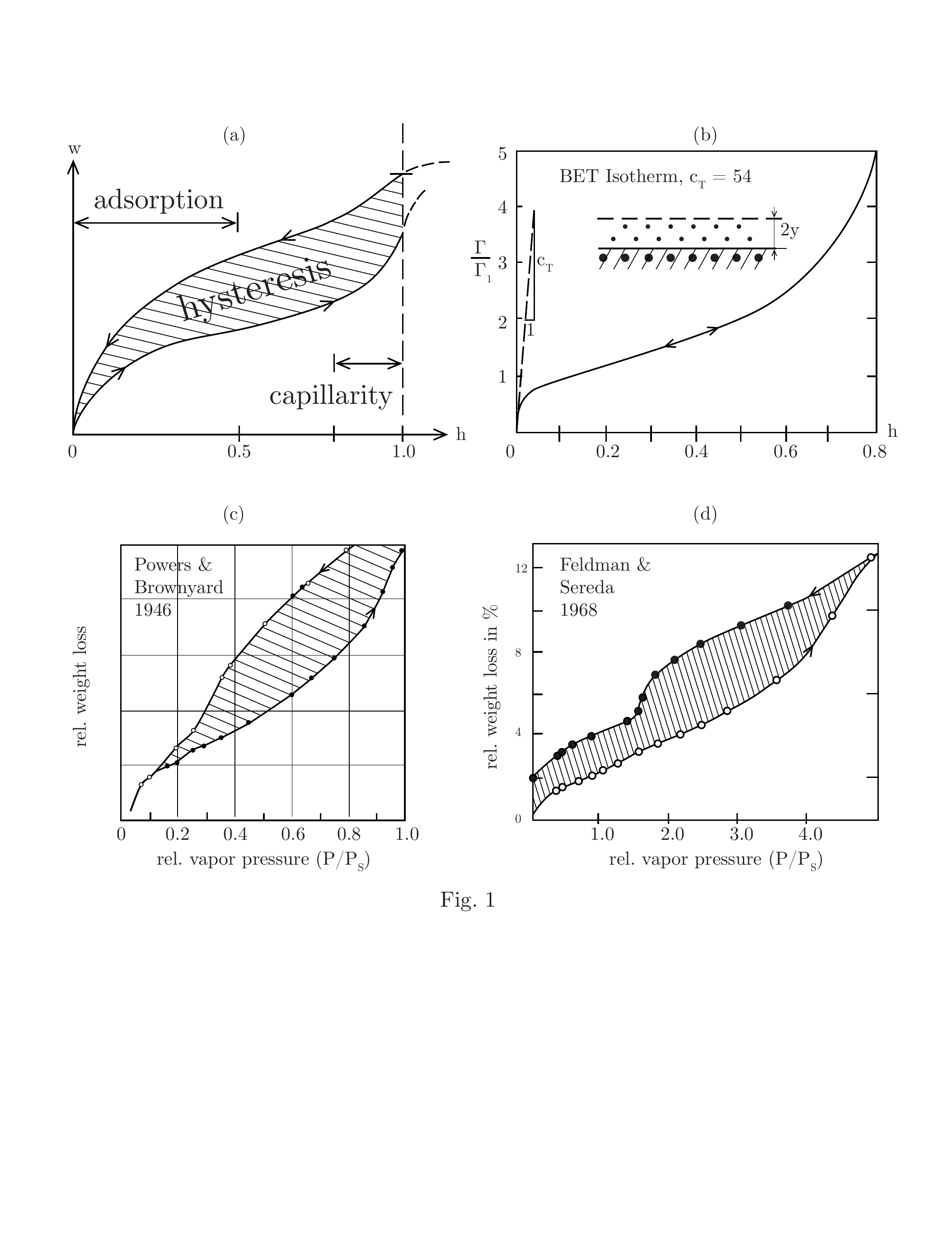}
\end{center}
\vspace{-2.5in}
\caption{ \label{1} \sf  (a) Typical desorption and sorption isotherms;
(b) BET isotherm; (c)-(d) Desorption and sorption isotherms measured on
hardened Portland cement paste. }
\end{figure*}

An important feature sorption experiments with water, nitrogen, alcohol,
methane, carbon dioxide, etc., has been a pronounced hysteresis, observed
at both high and low vapor pressures
and illustrated by two classical experiments in Fig. \ref{1}c,d) \cite[p.
277]{PowBro46} and \cite{FelSer64} (see also \cite[e.g.]{ FelSer68,
Rar-Jen95, AdoSet-02, EspFra06, baroghel2007}). For adsorbates that exist
at room temperature in a liquid form, e.g. water, the room temperature
hysteresis at high vapor pressures near saturation has easily been
explained by non-uniqueness of the surfaces of capillary menisci of liquid
adsorbate in larger pores (e.g., the `ink-bottle' effect \cite{Bru43}).
However, a liquid (capillary) water can exist in the pores only if the
capillary tension under the meniscus (which is given by the Kelvin-Laplace
equation) does not exceed the tensile strength of liquid water, which is
often thought to be exhausted at no less than 45\% of the saturation
pressure, if not much higher. Anyway, at vapor pressures less than about
80\% of the saturation pressure, the liquid phase represents a small
fraction of the total evaporable water content of calcium silicate hydrates
(C-S-H) \cite[Fig. 3]{Jen10}

However, the hysteresis at low vapor pressures (lower than 80\% of
saturation in the case of C-S-H) has remained a perplexing and unexplained
feature for over 60 years. In that case, most or all of the adsorbate is held
by surface adsorption. The gases and porous solids of interest generally
form adsorption layers consisting of several monomolecular layers (Fig.
\ref{1}b). The multi-layer adsorption is described by BET isotherm
\cite{BET38} (Fig. \ref{1}a), which is reversible. Sorption experiments
have generally been interpreted under the (tacit) hypothesis of free
adsorption, i.e., the adsorption in which the surface of the adsorption
layer is exposed to gas.

In nanoporous solids, though, most of the adsorbate is in the form of
hindered adsorption layers, i.e., layers confined in the nanopores (which
are sometimes defined as pores $\le 2$nm wide \cite{Bal-Gub93}). These
layers have no surface directly exposed to vapor and communicate with the
vapor in macropores by diffusion along the layer. It has been well known
that a large transverse stress, called the disjoining pressure \cite{Der40}
(or solvation pressure \cite{Bal-Gub93}), must develop in these layers.

Development of the theory of hindered adsorption for concrete was
stimulated by Powers' general ideas on the of creep mechanism \cite{Pow66}.
Its mathematical formulation for C-S-H gradually emerged in \cite{Baz70,
Baz70b, Baz72a, Baz72} and was reviewed in a broad context in \cite{Baz75}.
But this theory of hindered adsorption is also reversible. Thus, although a
theory exists, it cannot explain the hysteresis.

The sorption hysteresis in hardened Portland cement paste, concrete and
various solid gels \cite{Sch99, Jen-Sch08} has for a long time been vaguely
attributed to some sort of changes of the nanopore structure. In
particular, it was proposed that the exit or ingress of water, called the
interlayer water, from or into the narrowest nanopores would somehow cause
large relative changes of pore widths (\cite{FelSer68}; see also
\cite{Tho-Jen08}, Figs. 13 and 16 in \cite{Rar-Jen95}, Fig. 1 in
\cite{EspFra06}, or Fig. 9 in \cite{Jen10}), picturing pore width changes
$>100$\%. However, if a mathematical model of such a mechanism were
attempted it would inevitably predict enormous macroscopic deformations,
far larger than the observed shrinkage caused by drying. Here it will be
shown that sorption hysteresis must occur even if the nanopore structure
does not change.

The salient feature of capillarity is its non-uniqueness, and the main
message of this work is that an analogous non-uniqueness also applies to 
hindered adsorption in nanopores. In the capillary range, the
non-uniqueness is classically explained by the afore-mentioned
`'ink-bottle' effect, which exists even in two dimensions. In three
dimensions, there is much broader range of topological and geometrical
configurations which provide a much richer and more potent source on
non-uniqueness of liquid adsorbate content.

The simplest demonstration is a regular cubic array of identical spherical
particles separated by a small gap $\del$ between each pair.
At $h=1$, either all of the pore space can be filled by liquid, or an
anticlastic (hyperbolic paraboloid) meniscus surface of zero total
curvature $r^{-1} = r_1^{-1} + r_2^{-1} = 0$ and liquid pressure equal to
$p_s$ can exist between each two spheres, with $r_1 = - r_2$ (where $r_1,
r_2$ = principal curvature radii). This can explain 100\% differences among
equilibrium liquid contents $w$ at $h=1$ observed in some experiments. It
can even be shown that when both $\del$ and $r_1$ (with $r_2 = - r_1$)
approach zero in a certain way, then also the liquid content (as a
continuum) approaches 0. Thus, in theory, an arbitrarily small but nonzero
equilibrium liquid content at $h=1$ is possible, though extremely unlikely.

This three-dimensional picture, for example, explains why (as shown in Fig.
\ref{1}a by dashed lines) the non-uniqueness of sorption isotherm extends
to $h>1$ (where $h = p_v /p_s(T)$ = relative vapor pressure, or relative
humidity in the case of water, $p_v$ = pressure of vapor or gas; $p_s(T)$ =
saturation vapor pressure). For $h>1$, or $p_v > p_s$, the
total curvature of the menisci is changed from positive to negative, the
pores contain overpressurized vapor, and the hysteresis, or non-uniqueness
continues \cite{BazTho78, BazKap96}. This non-uniqueness and hysteresis
explains why the slope of the isotherm for $h>1$ is one, or even two,
orders of magnitude higher that one would calculate if all the water were
liquid for $h>1$. (In theory, this nonuniqueness can extend up to the critical
point of water). In cements these phenomena are complicated by the fact
that the chemical reactions of hydration withdraw some water from the
pores, and create self-desiccation bubbles. As a result, one practically never has
concrete devoid of any vapor, even for $p > p_s$. 

The consequence of the non-uniqueness is that the sorption isotherm is not
a function of local thermodynamic variables.  Instead, it is a functional of the entire previous history of adsorbate
content. Here we will show that the same functional character extends to the
range of hindered adsorption in nanopores, consistent with the extensive experimental data that consistently exhibits sorption hysteresis over the entire range of relative humidities.

The recent advent of molecular dynamic (MD) simulations is advancing the
knowledge of nanoporous solids and gels or colloidal systems in a profound
way \cite{ Pel-Ulm09, Coa-Pel09, Coa-Pel08, Coa-Pel08b, Jon-Wen05,
JonWen04, Smi-Whi06, Mal-Mur09, Van-Cou10}. Particularly exciting have been
the new results by Rolland Pellenq and co-workers at the Concrete
Sustainability Hub in MIT led by Franz-Josef Ulm \cite{Bon-Pel10,
Bro-Pel11, Bro-Pel11a}. These researchers used numerical MD simulations to
study sorption and desorption in nanopores of coal and calcium silicate
hydrates. Their MD simulations \cite[Fig. 3,4]{Bon-Pel10} demonstrated that
the filling and emptying of pores 1 and 2 nm wide by water molecules
exhibits marked hysteresis.

Especially revealing is the latest paper of Pellenq et al. from MIT
\cite{Bro-Pel11}. Simulating a chain of nanopores, they computed the
distributions of disjoining (or transverse) pressure and found that it can
alternate between negative (compressive) and positive (tensile), depending
on the difference of pore width from an integer multiple of the natural
thickness of an adsorbed monomolecular layer (see Figs. 4 and 11 in
\cite{Bro-Pel11}. This discrete aspect of disjoining pressure, which cannot
be captured by continuum thermodynamics, was a crucial finding of Pellenq
et al. which stimulated the mathematical formulation of snap-through
instabilities in Part I of this work. Oscillations between positive and
negative disjoining pressures have also been revealed by
density-functional-theory simulations of colloidal fluids or gels in
\cite{Bal-Gub93} (where the excess transverse stress is called the
``solvation pressure" rather than the disjoining pressure).

This work is organized as follows. In Part I, we begin by summarizing the
classical theory of multilayer adsorption on free surfaces by Brunauer,
Emmett and Teller (BET) ~\cite{BET38}, which is widely used to fit
experimental data, but assumes reversible adsorption without any
hysteresis. We then develop a general theory of hindered adsorption in
nanopores which accounts for crucial and previously neglected effects of
molecular discreteness as the pore width varies. This leads us to the first
general mechanism for sorption hysteresis, snap-through instability in
nonuniform pores, which is the focus of this Part I.

In Part II~\cite{part2}, we will show that attractive forces between discrete
adsorbed molecules can also lead to sorption hysteresis by molecular
coalescence in arbitrary nanopore geometries, including perfectly flat
surfaces and pores. This second mechanism for hysteresis is a general
thermodynamic instability of the homogeneous adsorbate that leads to stable
high-density and low-density phases below the critical temperature. The
mathematical formulation of the second part is thus based on
non-equilibrium statistical mechanics. Similar models have been developed for surface wetting by nanoscale thin films~\cite{degennes1985,gouin2008,gouin2009}, starting with Van der Waals over a century ago~\cite{vdw1893}.  Even more relevant models, accounting for nanoscale confinement, have been developed for ion intercalation in solid  nanoparticles with applications to Li-ion batteries~\cite{singh2008,burch2009,bai2011,dreyer2011}. In that setting, analogous phenomena of hysteresis~\cite{dreyer2010} (in the battery voltage vs. state of charge, in the limit of zero current) and nanoparticle size dependence~\cite{wagemaker2011} have now been observed in experiments. These connections, which convey the remarkable generality of hysteresis in adsorption phenomena,  will be developed more in the second part in the context of a statistical physics approach. Here, in the first part, we begin to build the theory using more familiar models from solid mechanics
and continuum thermodynamics.

\subsection*{Continuum Thermodynamics of Hindered Adsorption in Nanopores}

{\bf Free Adsorption:}\,
  When a multi-molecular adsorption layer on a solid adsorbent surface is in
contact with the gaseous phase of the adsorbate, the effective thickness
$a$ of the layer is well described by the BET equation \cite[eq.
28]{BET38}:   
 \beq \label{BET}
  \Theta  =\frac a {s_0} = \frac{\Ga_w}{\Ga_1}   
   = \frac 1 {1 - h} - \frac 1 {1 - h + c_T h},~~~~
   c_T = c_0 e^{\Del Q_a /RT}
 \eeq
where $T$ = absolute temperature; $\Ga_w$ = mass of adsorbate per unit
surface area; $\Ga_1$ = mass of one full molecular monolayer per unit area;
$\Theta=$ dimensionless surface coverage; $h$ = relative pressure of the
vapor in macropores with which the adsorbed water is in thermodynamic
equilibrium; $R$ = universal gas constant (8314 J kmole$^{-1}$
$^\circ$K$^{-1}$) ; $c_0$ = constant depending on the entropy of
adsorption; $\Del Q_a$ = latent heat of adsorption minus latent heat of
liquefaction; $s_0$ = effective thickness of a monomolecular layer of the
adsorbate; $a$ = effective thickness of the free adsorption layer (in
contact with vapor; Fig. \ref{1}b). For the typical value of $c_T$ = 54,
the BET isotherm is plotted in Fig. \ref{1}b, where the number of adsorbed
monolayers approaches five at the saturation pressure.

Eq. \ref{BET} can be easily inverted:
 \bea \label{inverse}
   h &=& h(a) = A + \sqrt{A^2 + B}          
 \\ \label{inverseA} \mbox{where}~~~~
  A &=& \frac{B c_T}2 \left(1 - \frac{s_0}a \right) ,~~~
  B = \frac 1 {c_T - 1}        
 \eea

{\bf Hindered Adsorption:}\,
  Consider now a pore with planar rigid adsorbent walls parallel to
coordinates $x$ and $z$ and a width $2y$ that is smaller than the combined
width $2a$ of the free adsorption layers at the opposite walls given by Eq.
(\ref{BET}). Then the adsorbate has no surface in contact with the vapor
and full free adsorption layers are prevented from building up at opposite
pore walls, i.e., the adsorption is hindered and a transverse pressure,
$p_d$, called the disjoining pressure \cite{Der40}, must develop. For water
in highly hydrophillic C-S-H,      
the adsorption layers can be up to 5 molecules thick, and so, in pores less
than 10 molecules wide ($2y < 2.6$~nm), hindered adsorption with disjoining
pressure will develop at high enough $h$. The adsorbent communicates by
diffusion of the adsorbate along the pore with the water vapor in an
adjacent macropore.

In a process in which thermodynamic equilibrium is maintained, the chemical
potentials $\mu$ of the vapor and its adsorbate, representing the Gibbs'
free energy per unit mass, must remain equal. So, under isothermal
conditions,
 \beq \label{dmu0}
  \dd \mu = \rho_a^{-1}(\dd \tilde p_d + 2 \dd p_a)/3 = \rho_v^{-1} \dd p_v
 \eeq
Here $\rho$ = mass density of the vapor and $\rho_a$ = average mass density
of the adsorbate (which probably is, in the case of water, somewhere
between the mass density $\rho_w$ of liquid water and ice). The superior
$\tilde{~}$ is attached to distinguish the disjoining pressure obtained by
continuum analysis from that obtained later by discrete molecular
considerations ($\tilde p_d = 0$ if the nanopore is not filled because the
transverse pressure due to water vapor is negligible); $p_a = \pi_a /y$ =
in-plane pressure in the adsorption layer averaged through the thickness of
the hindered adsorption layer; it has the dimension of N/m$^2$, and (in
contrast to stress) is taken positive for compression; $\pi_a$ =
longitudinal spreading 'pressure' in the adsorption half-layer of thickness
$y$ (here the term `pressure' is a historically rooted misnomer; its
dimension is not pressure, N/m$^2$, but force per unit length, N/m);
$\pi_a$ is superposed on the solid surface tension $\ ga_s$ which is
generally larger in magnitude, and so the total surface tension, $\ga =
\ga_s - p_a$, is actually tensile \cite[Fig. 2]{Baz72} (thus the decrease
of spreading pressure with decreasing $h$ causes an increase of surface
tension, which is one of the causes of shrinkage).

Further note that if $p_d$ and $p_a$ were equal, the left-hand side would
be $\dd \mu = \rho_a^{-1} \dd p_d$, which is the standard form for a bulk
fluid. Also, in contrast to solid mechanics, the left-hand side of Eq.
(\ref{dmu0}) cannot be written as $\eps_y \dd p_d + 2 \eps_x \dd p_a$
because strains $\eps_x$ and $\eps_y$ cannot be defined (since the
molecules in adsorption layers migrate and the difference between $p_d$ and
$p_a$ is caused by the forces from solid adsorbent wall rather than
by strains).

Consider now that the ideal gas equation $p_v \rho_v^{-1} = RT/M$ applies
to the vapor ($M$ = molecular weight of the adsorbate; e.g., for water $M$
= 18.02 kg/kmole). Upon substitution into Eq. (\ref{dmu0}), we have the
differential equation:
 \bea \label{dmu-}
  &&\mbox{for $h \le h_f$:}~~~~~\rho_a^{-1} \dd p_a = (RT/M)\ \dd p_v/p_v
 \\ \label{dmu}
  &&\mbox{for $h > h_f$:}~~~~~\rho_a^{-1}(\dd \tilde p_d + 2 \dd p_a) /3
                                 = (RT/M)\ \dd p_v/p_v
 \eea
where $h_f$ = value of $h$ at which the nanopore of width $2y$ gets filled,
i.e., $h_f = h(y)$ based on Eq. (\ref{inverse}). Factors 2 and 3 do not
appear for $h<h_f$ because the free adsorbed layer can expand freely in the
thickness direction. Integration of Eq. (\ref{dmu}) under the assumption of
constant $\rho_a$ yields:
 \bea \label{pa}
  && \mbox{for $h \le h_f$:}~~~~~~p_a  = \frac{\pi_a} y = \rho_a \frac{RT} M \ln h
 \\ \label{papd}
  && \mbox{for $h > h_f$:}~~~~~~\tilde p_d + 2( p_a - {p_a}_f )
     = 3 \rho_a \frac{RT} M \ln \frac h {h_f}
 \eea
where ${p_a}_f = p_a (h_f)$ = longitudinal pressure when the nanopore just
gets filled, i.e., when $a = y$.

It is now convenient to introduce the ratio of the increments of in-plane
and disjoining pressures,
 \beq \label{ka}
  \ka\; =\; \dd p_a\ /\ \dd \tilde p_d
 \eeq
which we will call the disjoining ratio. If the adsorbate were a fluid,
$\ka$ would equal 1. Since it is not, $\ka \ne 1$. The role of $\ka$ is
analogous to the Poisson ratio of elastic solids. A rigorous calculation of
$\ka$ would require introducing (aside from surface forces) the
constitutive equation relating $p_a$ and $p_d$ (this was done in
\cite{BazMos73}, but led to a complex hypothetical model with too many
unknown parameters).

We will consider $\ka$ as constant, partly for the sake of simplicity,
partly because (as clarified later) $\ka$ is determined by inclined forces
between the pairs of adsorbate molecules (Fig. \ref{3}b,c); $\ka$ should be
constant in multi-molecular layers because the orientation distribution of
these forces is probably independent of the nanopore width. Note that $\ka$
would equal 0 only if all these intermolecular forces were either in-plane
or orthogonal (Fig. \ref{3}a, as in a rectangular grid).

\begin{figure*}
\vspace{-1in}
\begin{center}
\includegraphics[width=6.5in]{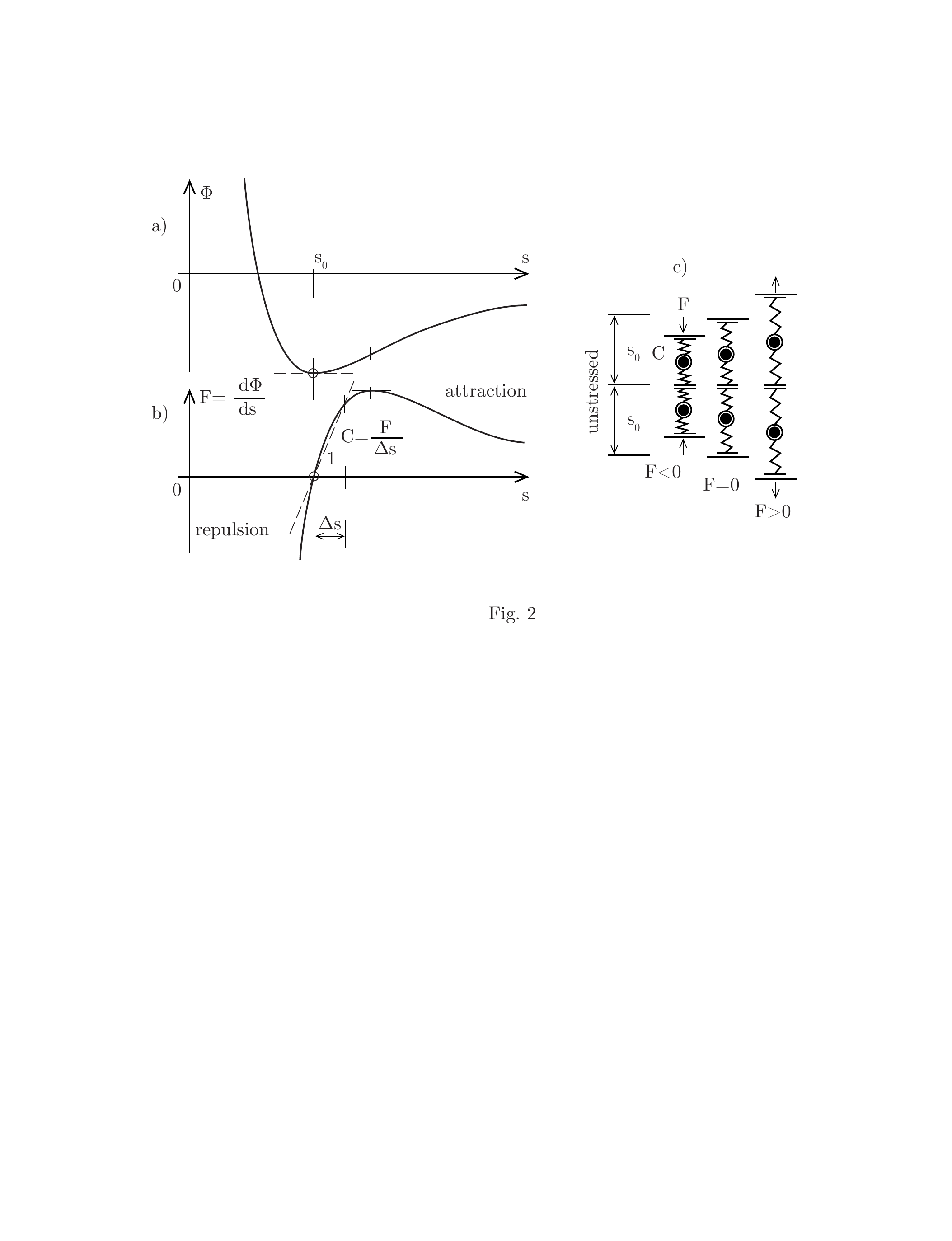}
\end{center}
\vspace{-4.5in} 
\caption{\label{2} \sf (a) Interatomic pair potential; (b) the
corresponding interatomic force and secant stiffness; (c) interatomic
forces between opposite pore walls visualized by springs.}
\end{figure*}

\begin{figure*}
\vspace{-1in}
\begin{center}
\includegraphics[width=6.5in]{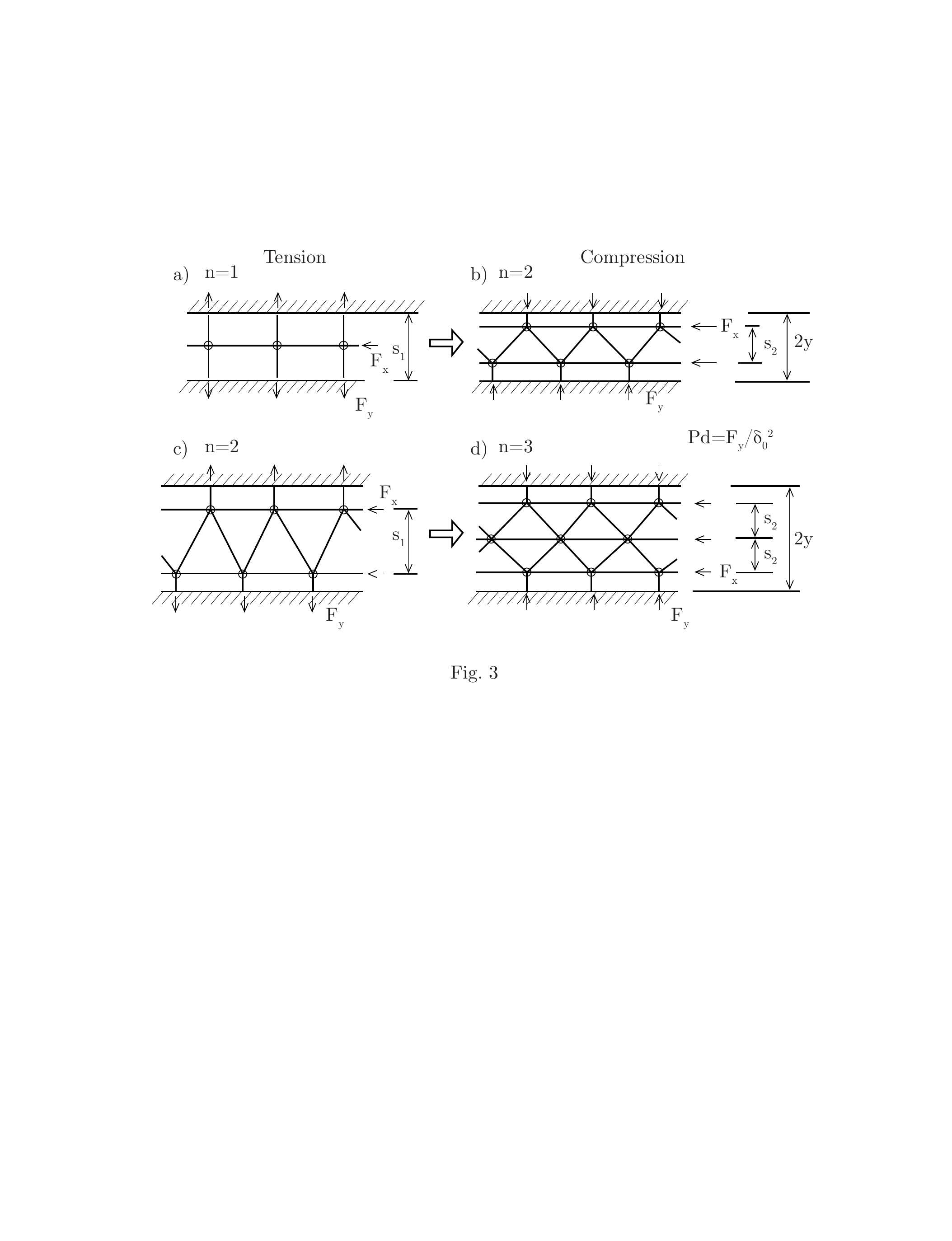}
\end{center}
\vspace{-4in}
\caption{\label{3} \sf Various simple idealized molecular arrangements
between the walls of a nanopore.}
\end{figure*}

For constant disjoining ratio $\ka$, we may substitute $p_a = \ka \tilde
p_d$ in Eq. (\ref{papd}), and we get
 \beq \label{pd}
  \tilde p_d\ =\ \frac{\rho_a}{1 + 2 \ka}\, \frac{RT}M\ \ln\frac h {h_f}
 \eeq
For $\ka$ = 0, this equation coincides with equation 29 in \cite{Baz72}
but, in view of Fig. \ref{3}, a zero $\ka$ must be an oversimplification.

According to this continuum model of hindered adsorption, which represents
a minor extension of \cite{Baz72}, the sorption isotherm of the adsorbate
mass as a function of vapor pressure would have to be reversible. However,
many classical and recent experiments \cite[e.g.]{PowBro46, FelSer64,
Rar-Jen95, EspFra06, baroghel2007} as well as recent molecular simulations
\cite{Bon-Pel10, Bro-Pel11, Bro-Pel11a} show it is not. Two mutually
related mechanisms that must cause sorption irreversibility in nanopores
with fixed rigid walls will be presented, one here in Part I, and one in
Part II which follows.

\subsection*{Mechanism I:  Snap-Through Instability  }

The local transverse (or disjoining) pressure $p_d$ can be determined from
the transverse stiffness $C_n$, defined as $C_n = \Del F /\Del s$ where
$\Del F$   
transverse resisting force per molecule and $\Del s$ = change of spacing
(or distance) between the adjacent monomolecular layers in a nanopore
containing $n$ monomolecular layers of the adsorbate. Since large changes
of molecular separation are considered, $C_n$ varies with $s$ and should be
interpreted as the secant modulus in the force-displacement diagram (Fig.
\ref{2}b). For this reason, and also because many bond forces are inclined
(``lateral interactions" \cite{Nik96, CerMed98, CerMed98b})

rather than orthogonal with respect to the adsorption layer (shown by the
bars in Fig. \ref{3}), $C_n$ is generally not the same as the second
derivative $\dd^2 \Phi /\dd r^2$ of interatomic potential nor the first
derivative $\dd F/\dd s$ of force $F$ (Fig. \ref{2}a,b).

To estimate $C_n$, one could consider various idealized arrangements of the
adsorbate molecules (as depicted two-dimensionally for two different pore
widths $2y$ in Fig. \ref{3}) and thus obtain analytical expressions for
$C_n$ based on the classical mechanics of statically indeterminate elastic
trusses. However, in view of all the approximations and idealizations it
makes no sense to delve into these details.


\begin{figure*}
\vspace{-1.5in}
\begin{center}
\includegraphics[width=6.5in]{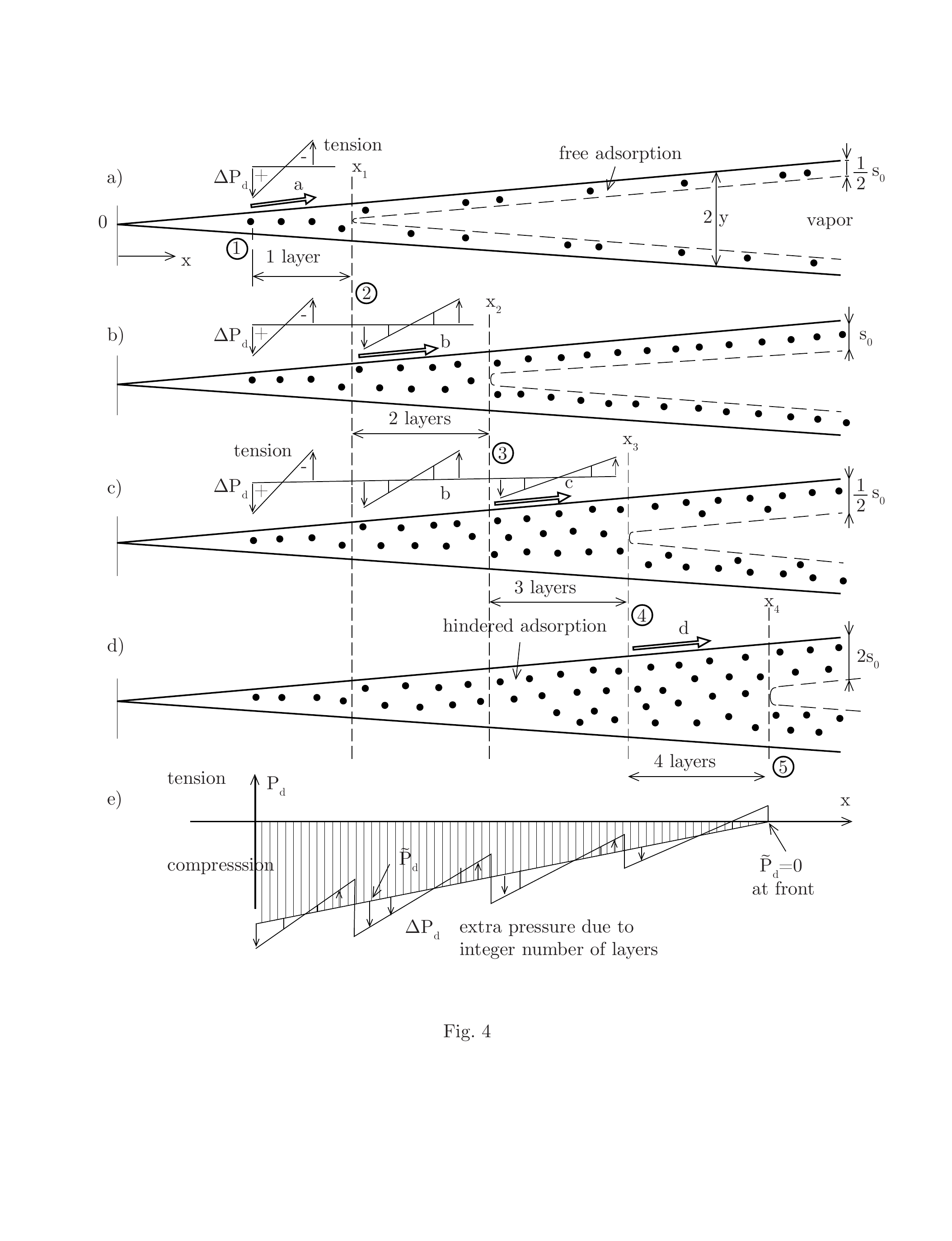}
\end{center}
\vspace{-1.5in}
\caption{\label{4} \sf Filling of a continuously diverging nanopore and
disjoining pressures.}
\end{figure*}

{\bf Diverging Nanopore:}\,
   Consider now a wedge-shaped nanopore between two diverging planar walls of
the adsorbent (Fig. \ref{4}a), having the width of $2y$ where
 $  y = kx $.
Here $x$ = longitudinal coordinate (Fig. \ref{4}), $k$ = constant (wedge
inclination) and $s_0$ = effective spacing of adsorbate molecules at no
stress. In the third dimension, the width is considered to be also $s_0$.
The adsorbate molecules are mobile and at the wide end (or mouth) of the
pore they communicate with an atmosphere of relative vapor pressure $h$ in
the macropores.

We assume the hindered adsorbed layer to be in thermodynamic equilibrium
with the vapor in an adjacent macropore. This requires equality of the
chemical potentials $\bar \mu$ {\em per molecule} ($\bar \mu = \mu /M$, the
overbar being used to label a quantity per molecule). At the front of the
portion of the nanopore filled by adsorbate, henceforth called the `filling
front' (marked by circled 2, 3 or 4 in Fig. \ref{4}), Eq. (\ref{pd}) of
continuum thermodynamics gives a zero transverse pressure, $\tilde p_d =
0$, and so $\bar \mu = \bar \mu_a = \bar \mu_v$. 

However, in the discrete treatment of individual molecules, the chemical
potential can be altered by transverse tension or compression $\Del p_d$
(Fig. \ref{4}), which can develop at the filling front and act across the
monomolecular layers unless the nanopore width $2y$ at the filling front
happens to be an integer multiple of the unstrained molecular spacing
$s_0$. We will call $\Del p_d$ the 'misfit' (part of) disjoining (or
transverse) pressure, by analogy with the misfit strain energy for a
dislocation core in the Peierls-Nabarro model~\cite{hirth}.

The misfit pressure, which, at the filling front, represents the total
transverse pressure (or stress), is determined by the average change $\Del
s$ of spacing $s$ between adjacent monomolecular layers, which is
 \beq \label{Deldel}
  \Del s = 2kx/n - s_0~~~~~~(n=1,2,3...)
 \eeq
where $n$ is the number of monomolecular layers across the nanopore width,
and $s_0$ is the natural spacing between the adjacent monomolecular layers
in free adsorption, i.e., when the transverse stress vanishes (note that
for the triangular arrangements in Fig. \ref{2}b,c, $s_0$ is obviously less
than the natural spacing of unstressed adsorbate molecules, shown as $s_0$
in Fig. \ref{2}a). So, the force between the molecules of the adjacent
layers is $F = C \Del s$ and the strain energy of the imagined springs
connecting the molecules is $F \Del s /2$ or $C (\Del s)^2 /2$ per molecule
(if, for simplicity, a loading along the secant is considered).

The hindered adsorbed layer is in a multiaxial stress state, for which the
total strain energy is the sum of the strain energies of the strain
components. Since continuum thermodynamics gives zero disjoining
(transverse) pressure $p_d$ at the filling front, it suffices to add to
$C_n (\Del s)^2 /2$ the chemical potential $\bar \mu_a$ per molecule at the
filling front due to longitudinal pressure $p_a$ only. So, in view of Eq.
(\ref{pd}), the chemical potential per molecule at the filling front
$x_n^f$ with $n$ monomolecular layers is
 \beq \label{mu n}
  {\bar \mu}_{f,n} =   
   \frac{C_n} 2 \left(\frac{2k x_f} n - s_0 \right)^2 + \bar \mu_a
 \eeq
where the overbar is a label for the quantities per molecule.
Since $\tilde p_d = 0$ at the filling front $x^*$, the only source of $\bar
\mu_n$ is the longitudinal spreading pressure $p_a$ in the adsorption
layer.

Let us now check whether at some filling front coordinate $x^*$ (Fig.
\ref{4}) the diverging nanopore is able to contain either $n$ or $n+1$
monomolecular layers with the same chemical potential per molecule. For
$n+1$ layers,
 \beq \label{mu n+1}
  \bar \mu_{f,n+1} =
   \frac{C_n} 2 \left(\frac{2k x}{n+1} - s_0 \right)^2 + \bar \mu_a
 \eeq
Setting $\bar \mu_n = \bar \mu_{n+1}$, we may solve for $x$. This yields
the critical coordinate and critical pore width for which the molecules in
$n$ and $n+1$ monomolecular layers have the same chemical potential per
molecule:
 \beq \label{xn}
  x_{f,n}^* = \frac{\sqrt{C_n} + \sqrt{C_{n+1}}}{\frac 1 n \sqrt{C_n}
        + \frac 1 {n+1} \sqrt{C_{n+1}}}\ \frac{s_0}{2k},~~~~~~y_{f,n}^* =
        2k x_{f,n}^*
 \eeq
So the critical relative pore width $2y_f^*/s_0$ at the filling front is a
weighted harmonic mean of $n$ and $n+1$ (and a simple harmonic mean if $C_n
= C_{n+1}$).

Equality of the chemical potentials per molecule at the filling front for
$n$ and $n+1$ monomolecular layers in the same nanopore, which occurs for
the pore width given by Eq. (\ref{xn}), implies that no energy needs to be
supplied and none to be withdrawn when the number of monomolecular layers
is changed between $n$ and $n+1$. So the equilibrium content of hindered
adsorbate in the nanopore for a given chemical potential of vapor is
non-unique. Similar to non-uniqueness of capillary surfaces, this
non-uniqueness underlies the sorption-desorption hysteresis in the
nanopores.

\begin{figure*}
\begin{center}
\includegraphics[width=6.5in]{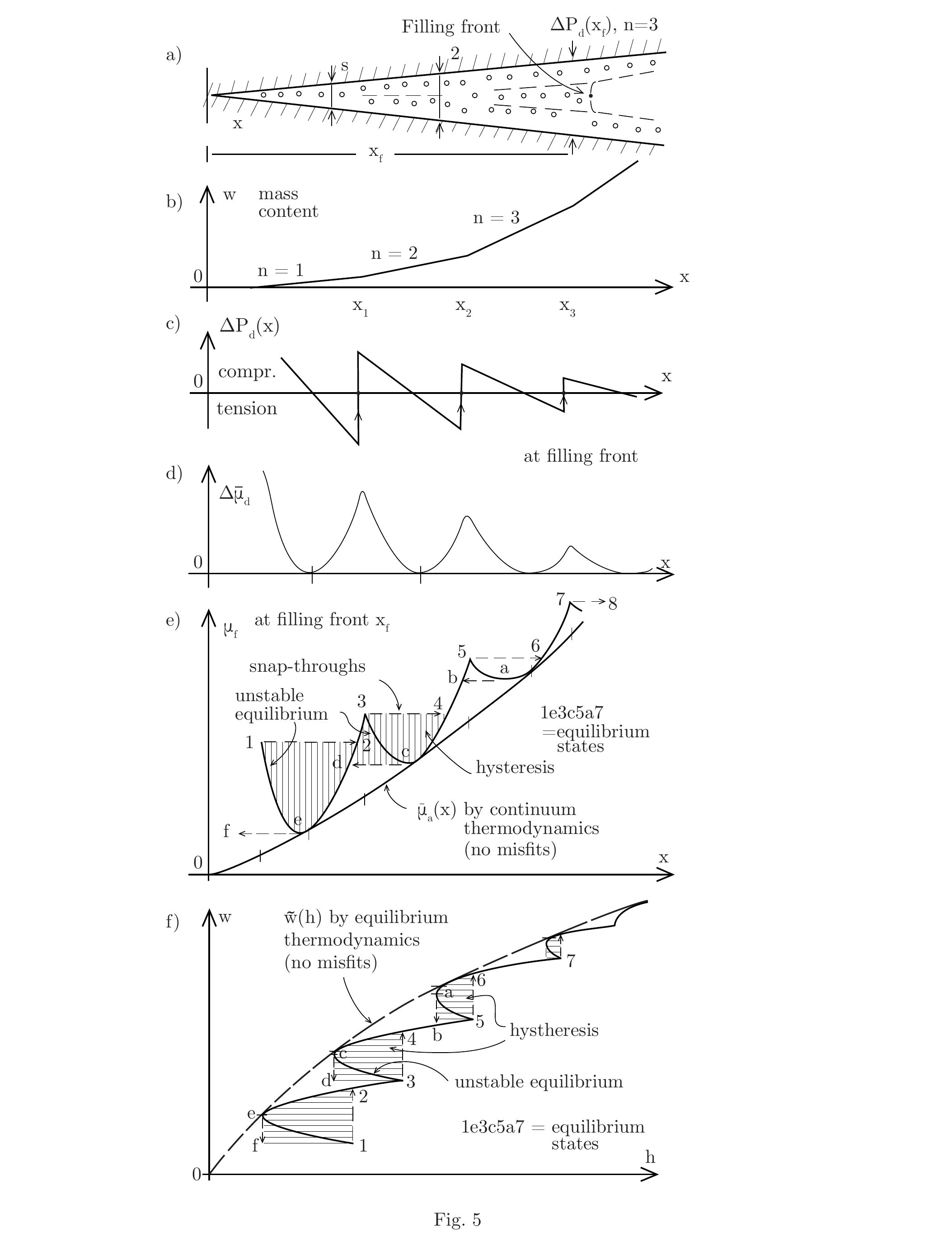}
\end{center}
\caption{\label{5} \sf Misfit disjoining pressures and chemical potentials
in a continuously diverging nanopore, with dynamic snap-throughs of
adsorbate content.}
\end{figure*}

{\bf Misfit Disjoining Pressure:}\, In view of Eq. (\ref{mu n}), its value
corresponding to $\bar \mu_n$ for $n$ monomolecular layers in the nanopore
is
 \beq \label{p n}
  p_{d,n} =    
  C_n \left( s_0 - \frac{2k x^*} n \right) + \tilde p_d(x_n)
 \eeq
where $\tilde p_d(x_n)$, based on continuum thermodynamics, is non-zero if
$x_n \ne x_{f,n}$. In contrast to stress, the pressure is considered as
positive when compressive. Replacing $n$ with $n+1$, we find that the
disjoining pressure makes a jump when the number of monomolecular layers in
the nanopore changes from $n$ to $n+1$;
 \beq \label{Del p}
  \Del p_{d,n} = p_{d, n+1} - p_{d,n} = 2k x_n \left( \frac{C_n} n -
    \frac{C_{n+1}}{n+1} \right) + s_0 \left(C_{n+1} - C_n \right)
 \eeq
(see Fig. \ref{4}). At the filling front, the jump is from transverse
tension to compression (Fig. \ref{5}c). The sudden jumps $\Del p_{d,n}$ of
the misfit pressures from tension to compression diminish with increasing
$n$ ($n=1,2,3,...$) as the wedge-shaped nanopore is getting wider; see
Figs. \ref{5}c and \ref{4}d. For $n > 10$, these jumps become
insignificant.

Note that, since the changes $\Del s$ of molecular distance are large, the
$C$ values depend on $F$ or $\Del p_d$ (Fig. \ref{2}b). So Eq. \ref{Del p}
is actually a nonlinear equation for $\Del p_d$ and its numerical solution
would require iterations. But here we are aiming at conceptual explanation
rather than numerical results.

{\bf Misfit Chemical Potentials and Their Effect on Sorption Isotherm:}\,
  The variation of chemical potential at the filling front $x_f$ is shown in
Fig. \ref{5}d. Since transverse tension at the filling front gives the same
chemical potential as transverse compression of equal magnitude, the misfit
chemical potential, defined as the part of chemical potential due to $p_d$
at the filling front, varies continuously, provided the pore width varies
continuously, too; see Fig. \ref{5}d. This is because transverse tension
gives the same chemical potential as transverse compression of equal
magnitude.

The total chemical potential at the filling front is obtained by adding the
chemical potential $\bar \mu_a (x_f)$ obtained from continuum
thermodynamics, which yields the potential variation in Fig. \ref{5}e.
Considering the relation of filling front coordinate $x_f$ to the adsorbate
mass $w$ shown (in a smoothed form) in Fig. \ref{5}b, and the relation $h =
e^{(M/RT) \mu_f}$,  
one can deduce the solid curve in Fig. \ref{5}e representing the diagram of
equilibrium states of mass content $w$ versus relative vapor pressure $h$
in the macropore.

Why are the segments of the pressure variation in Fig. \ref{5}c linear, and
why are the segments of the chemical potential variation in Fig.
\ref{5}d,e,f parabolic? The reason is that the variation of nanopore width
has been idealized as linear (and that the plots are made for constant
$C$). These segments take different shapes for other width variations.

\begin{figure}
\begin{center}
\includegraphics[width=6in]{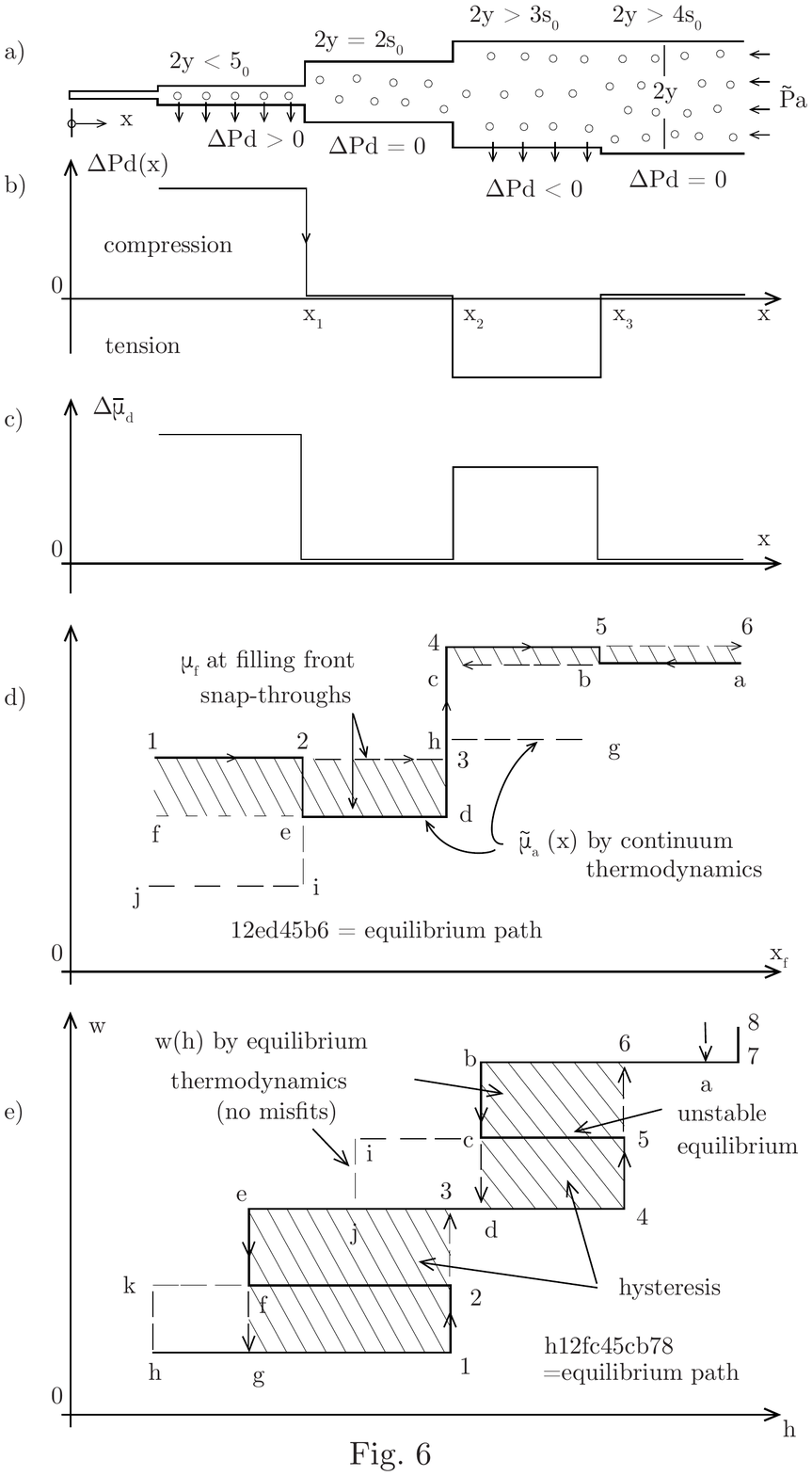}
\end{center}
\caption{\label{6} \sf Step-wise diverging nanopore and misfit disjoining
pressures.}
\end{figure}

{\bf Sequential Snap-Throughs of Adsorbate Content:}\,
    In sorption testing and most practical problems, the relative vapor
pressure $h$ is the variable that is controlled, and the adsorbate mass $w$
is the response. Consequently, the states at the reversal points 1, 3 5, 7
of the equilibrium diagram in Fig. \ref{5} for the diverging nanopore are
unstable. Likewise the states at points 1, 3, 5, 7 in Fig. \ref{6}d for the
nanopore of step-wise variable width. The loss of stability can be
evidenced by checking that the molecular potential loses positive
definiteness. Fundamental though such checks may be, it is simpler and more
intuitive to argue in terms of infinitely small deviations $\dd h$ from the
equilibrium state.

Consider, e.g., that, in Fig. \ref{5}f or \ref{6}d, a sufficiently slow
gradual increase of $h$ has moved the equilibrium state from point 2 to
point 3, which is a local maximum of $h$ as a function of $w$. For a
further infinitesimal increase $\dd h$ there is on the equilibrium diagram
no longer any point close to point 3. So, borrowing a term from
structural mechanics \cite{BazCed91}, we realize that the adsorbate mass
content $w$ must dynamically `snap through' at constant $h$ along vertical
line 34 to point 4. After dissipating the energy released along segment 34
(the rate of which depends on the lingering times of adsorbed molecules and
diffusion along the hindered adsorbed layer \cite{BazMos73}), thermodynamic
equilibrium is recovered at point 4. It is stable because a further
infinitesimal increment of $\dd h$ finds, next to point 4, an equilibrium
state with adsorbate content incremented by $\dd w$.

If $h$ is increased slowly enough further, the equilibrium system will move
from point 4 to point 5 at which a local maximum of $h$ is reached again
and the loss of stability gets repeated, since a further increase $\dd h$
can find equilibrium only after a dynamic snap-through to point 6. Each
snap-through will release some energy which must be damped and dissipated
by the system. So the local maxima of $h$ at points 1, 3, 5 and 7 are the
critical states giving rise to the so-called `snap-through instability'
\cite{BazCed91}. The equilibrium states on curved segments $1e2$, $3c4$ and
$5a6$ are unstable and can never be reached in reality.

The salient feature is that a different path $w(h)$ is followed when $h$ is
decreased. To show it, consider point 7 in Fig. \ref{5}f or \ref{6}d as the
starting point. During a slow enough decrease $h$, the system will follow
the stable states along segment $76a$ until a local minimum of $h$ is
reached at point $a$, which is the stability limit. Indeed, if $h$ is
further decremented by $\dd h$, there is no equilibrium state near point
$a$. So the equilibrium state $a$ is unstable and the system will `snap
through' dynamically at constant $h$ along path $ab$. At point $b$ stable
equilibrium is regained after sufficient time. When $h$ is decreased
further slowly enough, the equilibrium states move through segment $b4c$
until again a local minimum of $h$ is reached and stability is lost at
point $c$. Thereafter, the system `snaps through' along line $cd$ to point
$d$, where equilibrium is regained, etc.

In the diverging pore in Fig. \ref{5}, the snap-through means that when the
equilibrium filling front reaches the critical points, $x_1, x_2$, or $x_3$,
it will advance forward a certain distance at constant $h$, as fast as
diffusion along the micropore, controlled by the lingering times of the
adsorbate molecules, will permit.
The cross-hatched areas in between the sorption and desorption isotherms,
such as area $34cd3$ in Fig. \ref{5}e or Fig. \ref{6}d, represent sorption
hysteresis. They also characterize energy dissipation.

{\bf Sequential Snap-Throughs for Step-Wise Nanopore Width Variation:}\,
  The diagrams in Fig. \ref{5}d,e,f are valid only for a micropore with
continuously diverging rigid planar walls (Fig. \ref{5}a). This is, of
course, an idealization. Because of the atomistic structure of pore walls,
the pore width in reality varies discontinuously, as exemplified in Fig.
\ref{6}a. The chance of a width exactly equal to an integer multiple of
$s_0$ is small.

Consider that the jumps of nanopore width (Fig. \ref{6}a) occur at $x_1,
x_2, x_3,...$, and that at $x_1$ is narrower than $s_0$, at $x_2$ exactly
equal to $2 s_0$, and at $x_3$ wider than $3 s_0$. Thus the filling front
in pore segment $(x_1,x_2)$ is in transverse compression, in segment $(x_2,
x_3)$ at zero transverse pressure, and in segment $(x_3, x_4)$ in
transverse tension; see Fig. \ref{6}b. The corresponding strain energies,
representing the misfit chemical potential $\Del \bar \mu_d$ per molecule,
have a pulse-like variation as shown in Fig. \ref{6}c. Continuum
thermodynamics, which ignores the misfits, gives a monotonically rising
staircase variation of the chemical potential $\tilde \mu_a(x)$ (per unit
mass) as a function of the filling front coordinate $x_f$, represented by
path $jiedhgba$ (Fig. \ref{6}d). Superposing on this staircase the misfit
chemical potential $\Del \bar \mu_d$ (converted to unit mass), one gets the
non-monotonic step-wise path of equilibrium states, shown by the bold line
12ed455b6 in Fig. \ref{6}d.

Taking into account the dependence of the adsorbate mass $w$ in the
nanopore on the filling front coordinate $x_f$, one can covert the diagram
in Fig. \ref{6}d into the sorption isotherm in Fig. \ref{6}e, usually
plotted as $w$ versus $h$. The monotonic staircase $hkfejicba$ would
represents the equilibrium path if the misfit disjoining pressures were
ignored.

When the rise of $h$, and thus $\mu_f$, is controlled, the segments 23 and
56 in Fig. \ref{6}e are unstable and unreachable. Indeed, when $h$ or
$\mu_f$ is infinitesimally increased above point 2, there is no nearby
equilibrium state, and so the system will `snap through' dynamically to
point $3$. At that point, equilibrium is regained, and $h$ and $\mu_f$ can
be raised again, slowly enough to maintain equilibrium, along path 345. A
similar dynamic snap-through is repeated along segment 56, after which the
stable segment 678 can be followed. Likewise, in the diagram of $\mu_f$
versus the filling front coordinate $x_f$ (Fig. \ref{6}d), forward
snap-throughs at increasing $\mu_f$ (which is a monotonically increasing
function of $h$) occur along segments 23 and 56.

When $h$ or $\mu_f$ is decreased slowly enough from point 8, the stable
equilibrium path $876bc$ is followed until stability is lost at point $c$
(Fig. \ref{6}e). Then the system snaps through dynamically from $c$ to $d$,
follows equilibrium path $def$, and snaps dynamically from $f$ to $g$.
Likewise, in Fig. \ref{6}d, backward snap-throughs at decreasing $\mu_f$
occur along segments $bc$ and $ef$.

Obviously, the states on segments $c5$ and $f2$ in Fig. \ref{6}e, or $2e$
and $5e$ in Fig. \ref{6}d, can never be reached. They represent unstable
equilibrium. The shaded areas $g13eg$ and $d46bd$ represent hysteresis,
which leads to energy dissipation.

{\bf Snap-Throughs in a System of Nanopores:}\,
   The diverging nanopore (Fig. \ref{4}, \ref{5}a and \ref{6}a) is not the
only pore geometry producing sorption hysteresis. There are infinitely many
such geometries. In the simple model of discrete monolayers pursued in Part
I, the only geometry avoiding hysteresis due to sequential snap-throughs is
hypothetical---the widths of all the nanopores would have to be exactly
equal to the integer multiples of the natural spacing $s_0$ of
monomolecular layers in free adsorption, so as to annul the misfit
pressures. Below, we will show that if molecular coalescence is allowed in
the lateral direction, then even these special pore geometries will exhibit
sorption hysteresis, and so the effect is extremely general.

\begin{figure*}
\begin{center}
\includegraphics[width=6.5in]{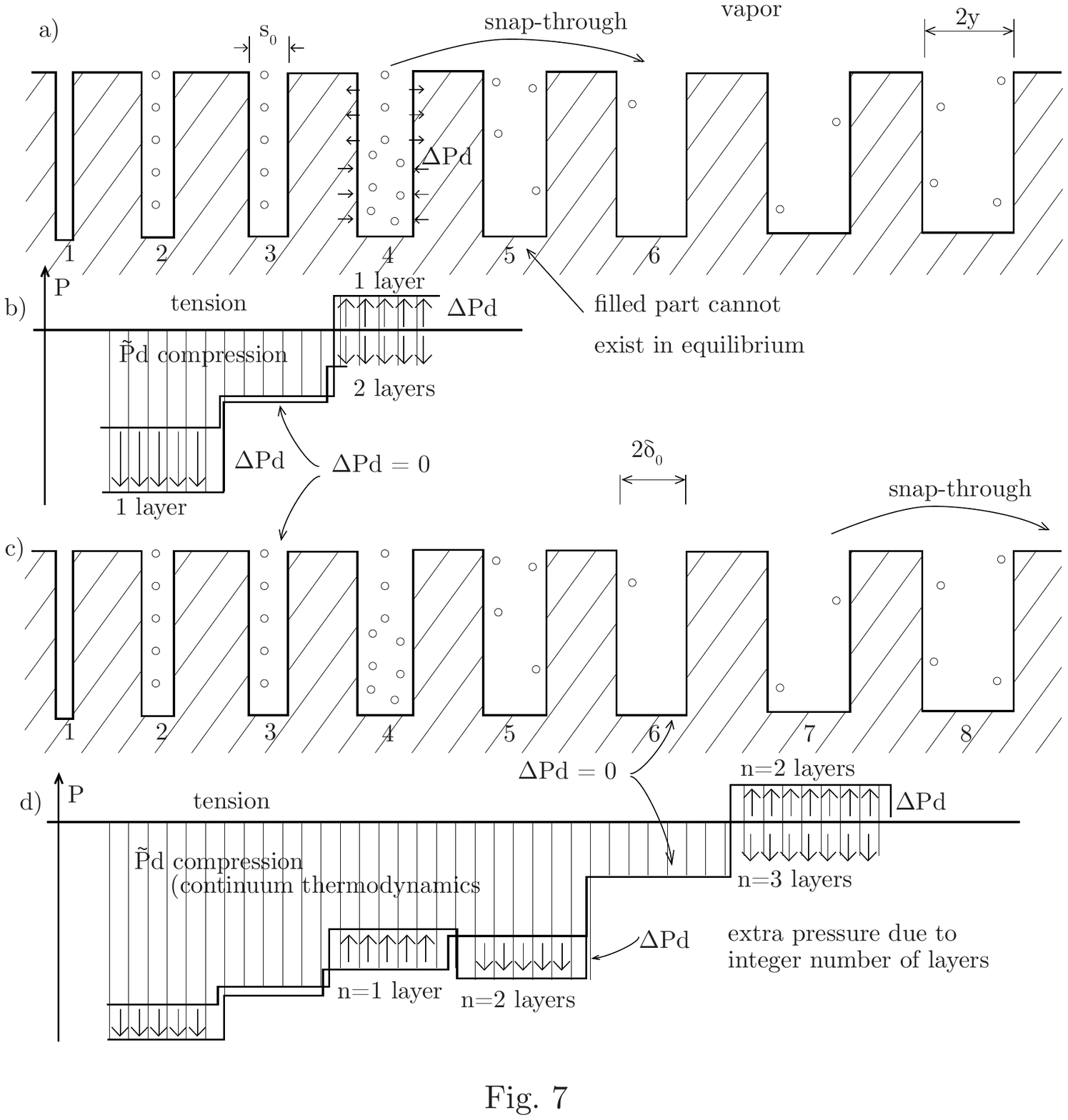}
\end{center}
\vspace{-1.7in}
\caption{\label{7} \sf System of nanopores of different widths
communicating through vapor phase .}
\end{figure*}

An essential feature of nanoporosity is that there are nanopores of many
different thicknesses $2y$ densely distributed as shown in Fig. \ref{7}. At
a given vapor pressure, all the nanopores that are narrower than a certain
width $2y$ are filled by adsorbed water and the wider ones are empty,
containing only vapor; see Fig. \ref{7}a,c,e.

As the relative pore pressure $h$ is increased, larger and larger pores
fill up. A critical state (or a local maximum of $h$) is reached for a pore
width at which the misfit chemical potential $\Del \mu_d$ due to misfit
disjoining pressure is for $n$ monomolecular layers equal to or larger than
the misfit chemical potential for $n+1$ layers. After that state, the
system loses stability and regains it only when all the nanopores up to a
certain larger width get filled without increasing $h$. For decreasing $h$,
the stability loss would occur for a different pore width.

The distribution of nanopore thicknesses $2y$ may be characterized by a
continuous cumulative frequency distribution function $\varphi(y)$ that
represents the combined volume of all the nanopores with thicknesses $<
2y$. This case, though, is not qualitatively different from the diverging
nanopore studied previously. For $\varphi(y) \propto k y^2$, the nanopore
system in Fig. \ref{7} becomes mathematically equivalent to the linearly
diverging nanopore studied before.

\begin{figure*}
\vspace{-1in}
\begin{center}
\includegraphics[width=6.5in]{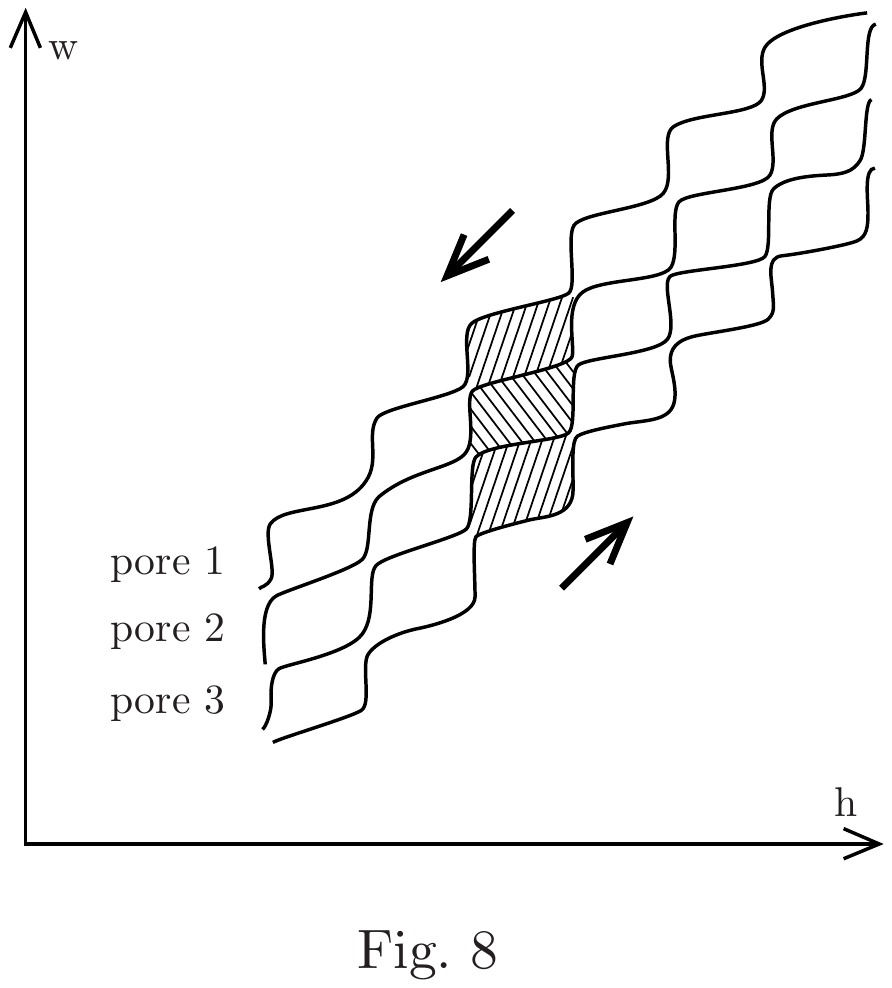}
\end{center}
\vspace{-4.5in}
\caption{\label{8} \sf Superposition of hysteretic loops from different
nanopores.}
\end{figure*}

The way the hysteresis in the individual nanopores gets superposed to
produce a pronounced hysteresis on the macroscale is schematically
illustrated in Fig. \ref{8}.

\nopagebreak
\begin{figure*}
\vspace{-1in}
\begin{center}
\includegraphics[width=7in]{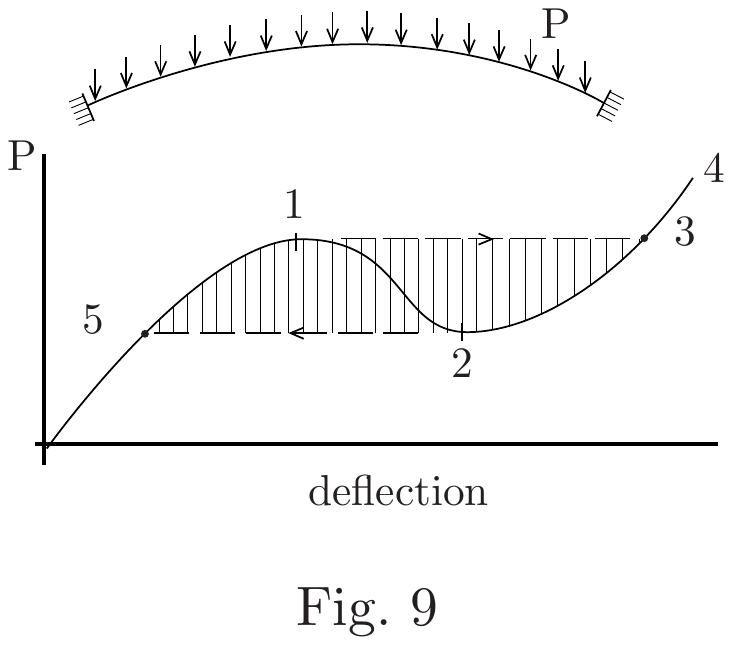}
\vspace{-5.5in}
\caption{\label{9} \sf Analogy with snap-through of an arch.}
\end{center}
\end{figure*}

{\bf Analogy with Snap-Through Buckling of Flat Arch:}\,
   There is an instructive analogy with the snap-through buckling of elastic
arches or shells under controlled load (Fig. \ref{9}
\cite[p.231]{BazCed91}. If the arch is flat enough and flexible enough not
to fracture, the equilibrium diagram of total load $p$ versus midspan
deflection $u$ follows the diagram sketched in Fig. \ref{9}. The segments
051 and 432 consist of stable states at which the potential energy is
positive definite (i.e., has a strict local minimum). But this is not true
for the equilibrium states on the segment 12, at which the potential energy
does not have a strict local minimum.

Consider that load $p$ is increased from point 0 up to the local maximum at
critical point 1 (Fig. \ref{9}). If load p is increased further by an
infinitesimal amount $\ dd p$, there is no nearby equilibrium state. The
arch must follow at constant load the dynamic snap-through path 14, during
which there is accelerated motion, with the load difference from the
equilibrium curve below being equal to the inertial force, which provides
rightward acceleration. The arch gains kinetic energy up to point 3, swings
over (along a horizontal line), and then vibrates at constant load about
point 3 until the kinetic energy is dissipated by damping (without damping,
it would vibrate indefinitely). Then, if the load is increased further, the
arch moves through stable equilibrium states on the segment 34.

When the load is decreased, starting at point 4, the arch will follow the
stable equilibrium states along segment 432 until a local minimum is
reached at point 2. If the load is decreased further by an infinitesimal
amount $\dd P$, there is no equilibrium state near point 2. So the arch
must snap through dynamically to point 5, the load being again balanced by
inertia forces which provide leftward acceleration. During this
snap-through the arch gains kinetic energy, swings over ot the left of
point 5 and vibrates about point 5 until the kinetic energy is dissipated
by damping. Then the load can be decreased further following the stable
equilibrium states below point 5.

Note that even though the arch is elastic and the structure-load system is
conservative, hysteresis is inevitable. During the cycle, the arch
dissipates an energy equal to the cross-hatched area 51325 in Fig. \ref{4}.


{\bf Energy Dissipated by Hysteresis and Material Damage:}\,
   The Gibbs free energy dissipated per unit mass of the nanoporous material is
$\dd G = w\, \dd \mu$ where $\dd \mu = \frac{RT}{M} \, \dd \ln h$, 
which has
in thermodynamic equilibrium the value for the adsorbate species in the
vapor and for the adsorbed phases. Therefore, the free energy dissipated
per unit volume of material due to the hysteresis during a complete cycle,
e.g., a drying-wetting cycle of hardened cement paste, is
 \beq  \label{dissip}
  \Del G = \frac{RT} M \oint \frac{w(h)} h \dd h
 \eeq
Since $h$ is in the denominator, integrability, i.e., the finiteness of
$\Del G$, requires that $\lim_{h \to 0} w/h = 1/h^n$ where $0 \le n < 1$.
Graphically, $\Del G$ is proportional to the area between the sorption and
desorption isotherms in the diagram of $w/h$ versus $h$ (Fig. \ref{dissip
en}).

\begin{figure*}
\begin{center}
\vskip -.5in
\includegraphics[width=5.9in]{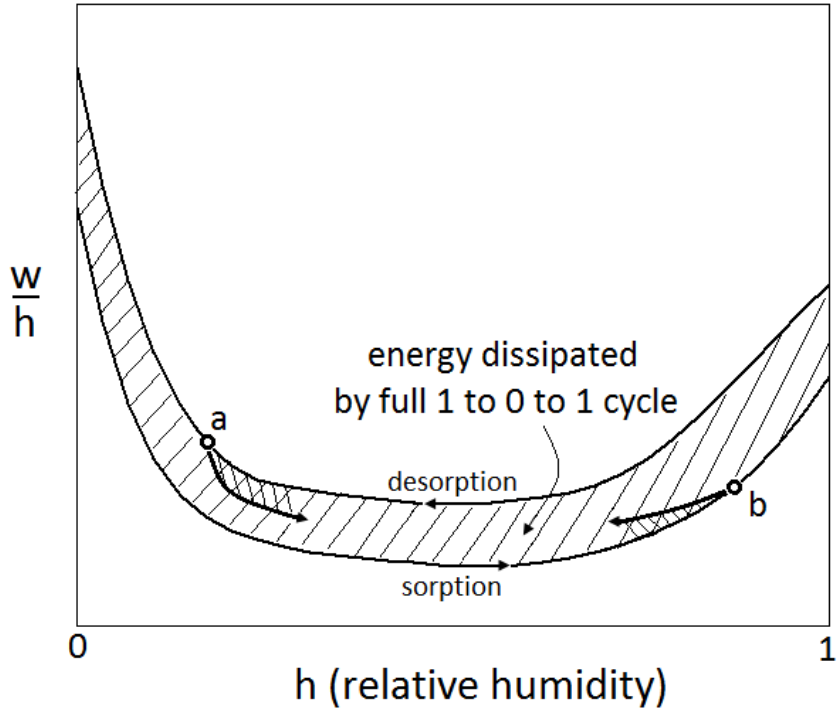}
\vspace{-.1in}
\caption{\label{dissip en} \sf Energy dissipated by sorption hysteresis on
a full $h$-cycle $1 \to 0 \to 1$ (shaded area), and dissipation during
mid-range reversals (a,b).}
\end{center}
\end{figure*}

The energy could be dissipated in two ways:
\\ \hhh 1) By internal friction in the adsorbed fluid during the
dynamic snap-throughs (or the molecular coalescence phenomena discussed in Part II~\cite{part2}), or
\\ \hhh 2) by fracturing or plastic damage to the nanopore surfaces.
\\ However, the latter seems unlikely since it could be associated with
every disjoining pressure change and not particularly with the
snap-through. The existence of the former is undeniable, and the point here
is to show that the hysteresis is perfectly explicable without postulating
any damage to the nanopore surface.

Anyway, the degree of material damage due to a drying-wetting cycle, if
any, could be checked by measuring the strength or the fracture energy, or
both, of the material before and after the cycle. This would have to be
done slowly enough on thin enough specimens having drying half-times less
than 1 hour ($<1$ mm thick for cement paste), in which the relative
humidity $h$ in the capillary pores can be changed without creating a
significant gradient of $h$ across the specimen wall (in thicker samples,
most of the material damage is done by non-uniform shrinkage stresses
engendered by non-uniformity of $h$ across the wall thickness
\cite{BazRaf82}). Shrinkage and creep experiments on such specimens have
been performed at Northwestern \cite{BazAsg-76}, but no cycles were
performed and strength changes were not checked. It could also be checked
whether the snap-throughs might be associated with the acceleration of
concrete creep due to simultaneous drying, called the drying creep (or
Pickett effect).

{\bf Sorption Potential:}\,
  Note that, based on the derivation of Eq. (\ref{dissip}), it further
follows that
 \beq \label{potential}
  {\beta} \; = \; \frac {RT}M \, \frac w h \; =\; \frac{\pa G}{\pa h}
 \eeq   
In other words, the Gibbs's free energy per unit mass of adsorbate as a
function of $h$ is a potential for the adsorbate content parameter
${\beta}$ during a one-way change of $h$.

\subsection*{Conclusions of Part I }

We can summarize the findings of the first part as follows:
 \be \ii
One mechanism that must be causing sorption hysteresis at low vapor
pressure is a series of snap-through instabilities causing path-dependent
non-uniqueness of adsorbate content and dynamic jumps of water content of
nanopores at constant vapor pressure.
 \ii
The snap-through instabilities are a consequence of the discreteness of the
adsorbate, which leads to non-uniqueness of mass content and to misfit
disjoining (transverse) pressures due to a difference between the pore
width and an integer multiple of the thickness of a transversely unstressed
monomolecular layer of the adsorbate.
 \ii
The hysteresis is explained by the fact that the snap-through instabilities
for sorption and desorption follow different paths.
 \ii
The snap-through instabilities are analogous to snap-through buckling of
arches and shells, long known in structural mechanics. They cause
hysteresis and energy dissipation even when the arch or shell is perfectly
elastic.
 \ee
If a quantitative version of this theory were developed, it might be
possible to infer from the hysteresis the surface area and the size
distribution of the nanopores filled by hindered adsorbate. Our preliminary
analysis of snap-through instabilities suggests that the key to making this
connection is to account for inclined forces, or ``lateral interactions",
in the statistical thermodynamics of hindered adsorption. In the Part II,
we will show that attractive lateral interactions generally lead to
sorption hysteresis in any pore geometry due to molecular coalescence of
the adsorbate.

 {\small {\sf
\vv \no {\bf Acknowledgment:} The research was funded partly by the U.S.
National Science Foundation under Grant CMS-0556323 to Northwestern
University (ZPB) and Grant DMS-0948071 to MIT (MZB) and partly by the U.S.
Department of Transportation under Grant 27323  provided through the
Infrastructure Technology Institute of Northwestern University (ZPB).
Thanks are due to Franz-Josef Ulm and Rolland J.-M. Pellenq of MIT for
stimulating discussions of disjoining pressure based on MD simulations, and
to Laurent Brochard and Hamlin M. Jennings for further valuable discourse.
 }}


\end{document}